\documentclass[]{aastex631}
\usepackage{amsmath,amssymb}   
\usepackage{graphicx}          
\usepackage{url}
\usepackage{color}

\newcommand{\mb}[1]{\boldsymbol{#1}}     
\newcommand{\unit}[1]{\;\mathrm{#1}}        
\mathchardef\mhyphen="2D                 
\shorttitle{Temporal Evolution of Europa's Ice Shell}
\shortauthors{Shibley \& Goodman}
\graphicspath{{./}{figures/}}

\begin{document}

\title{Europa's Coupled Ice-Ocean System: Temporal Evolution of a Pure Ice Shell}

\correspondingauthor{Nicole C. Shibley}
\email{nicole.shibley@princeton.edu, jeremy@astro.princeton.edu}

\author{Nicole C. Shibley}
\affiliation{Princeton Center for Theoretical Science \\
Princeton University \\
Princeton, NJ 08544 USA}

\author{Jeremy Goodman}
\affiliation{Princeton Center for Theoretical Science \\
Princeton University \\
Princeton, NJ 08544 USA}
\affiliation{Department of Astrophysical Sciences \\
Princeton University \\
Princeton, NJ 08544, USA}



\begin{abstract}
Europa's icy surface likely overlies an ocean, but the ice thickness is not known. 
Here we model the temporal growth of a Europan shell of pure ice subject to varying ice-ocean heat fluxes, ice rheologies, and internal heating rates. Both constant and viscosity-dependent internal heating rates are included, yielding similar results for particular viscosities. A growing shell starting from an ice-free initial state transitions from conduction to convection at O(10$^5$) to O(10$^7$) years, with thicknesses O(1-10)~km. For low ice-ocean heat fluxes and larger viscosities, the time to reach a steady-state thickness exceeds the estimated age of Europa's surface, whence the shell may still be growing. We conclude by presenting a method for inferring ice-ocean heat fluxes and vertical ocean velocities from the ice-thickness measurements expected from the upcoming \textit{Clipper} mission, assuming the shell is in a conductive steady state.
\end{abstract}



\section{Introduction} \label{sec:intro}
Europa is an ice-covered moon of Jupiter with a purported subsurface ocean \cite[e.g.,][]{carr1998, pappalardo1999}. Given the likely presence of liquid water, much interest has been generated over the potential for astrobiology in Europa’s ocean-ice system \citep[e.g.,][]{hand2009}, and the optimal places to find it \citep{buffo2021, steinbrugge2020}. However, little is known about the coupled ice-ocean system. Given that the ice shell is a primary observable of Europa and is a surface manifestation of its subsurface ocean, understanding the mechanisms which govern the shell growth may give insight into the ocean below. 

Several zeroth-order questions about Europa’s ice cover remain. In particular, the thickness of Europa’s ice cover is not well constrained, with estimates spanning two orders of magnitude \citep{ojakangas1989, hoppa1999, mckinnon1999, turtle2001, quick2015, howell2021}. Geological investigations focusing on surface morphology give conflicting results. Cycloidal cracks along Europa's surface suggest a thin ice cover, generally O(1)~km \citep[e.g.,][]{hoppa1999, greenberg2000}. However, certain features of Europa's terrain have been interpreted as convective diapirs, which can only occur in a thick shell, generally of O(10)~km \citep[e.g.,][]{pappalardo1998}.

The thickness of Europa’s ice shell is necessarily tied to the mechanisms of heat transport through it. The upper surface of the shell is exposed to space and thus is extremely cold, around 100~K or so depending on the latitude \citep[e.g.,][]{ashkenazy2019}. However, the ice-ocean interface of Europa’s ice shell is tied to the freezing temperature of water, approximately 273~K (with modest corrections for salinity and pressure). Thus, a steep temperature gradient exists across the ice shell. This means that heat wants to move across the shell from the basal ice-ocean boundary towards the ice shell surface, where it is lost radiatively to space.  

\begin{figure}
    \includegraphics[width=\textwidth]{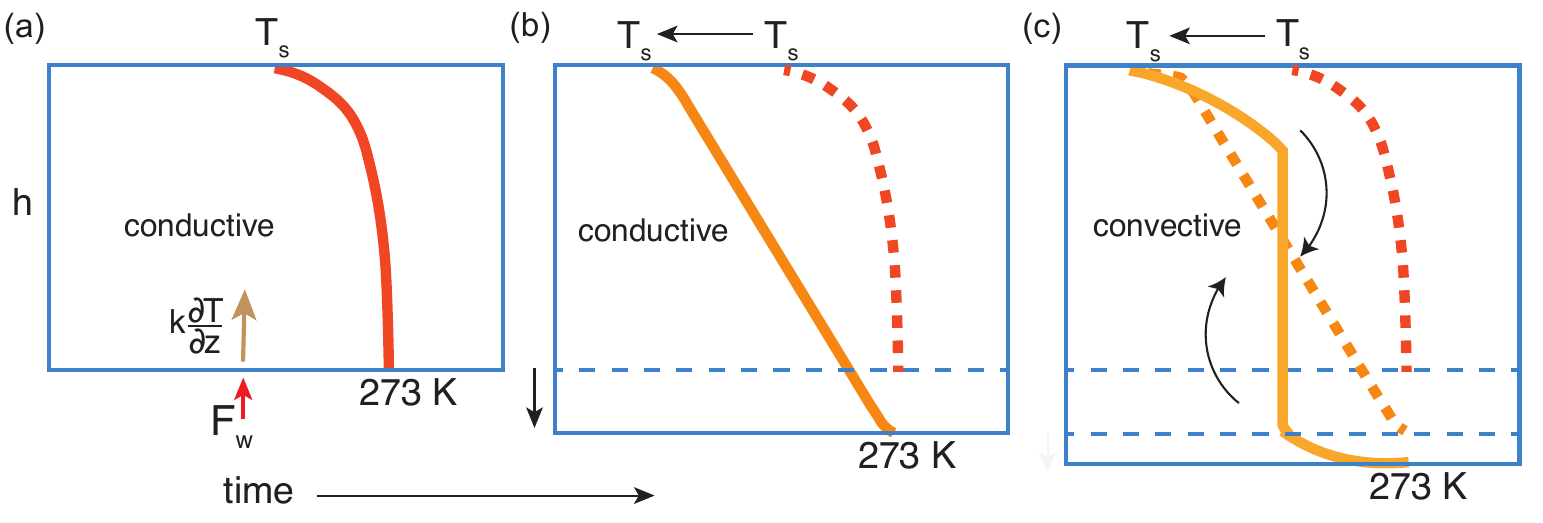}
        \caption{(a) A temperature profile indicative of conductive transfer across the shell. The surface is at a temperature $T_s$, while the basal temperature is tied to the freezing point of water. The shell experiences an ice-ocean heat flux $F_w$ at its base. (b) As the shell thickens, the surface temperature decreases. If $Ra < Ra_c$, heat transport occurs by conduction. (c) Once the shell thickens sufficiently, the critical Rayleigh number $Ra_c$ can be exceeded, and the shell becomes unstable to convection. Then, there is an almost constant temperature profile sandwiched between two boundary layers.}\label{fig1}
\end{figure}

This vertical heat transport may occur one of two ways: conduction or convection. For heat to be transported via molecular conduction alone, the ice shell must be thin. Otherwise, the shell can become unstable to convection, which can transport heat more rapidly than a molecular process \citep{benard1900, rayleigh1916} (Figure \ref{fig1}). However, convective motions are resisted by viscosity, while the horizontal temperature variations that drive these motions via thermal expansion are smoothed by conduction.
The competition among these effects is often described by the Rayleigh number, $Ra$, defined as:
\begin{equation}
    Ra\equiv\frac{\rho g\beta\Delta TH^3}{\eta\kappa},
\end{equation}
where $\rho$ is the density of the fluid, $g$ is gravity, $\beta$ is the coefficient of thermal expansion, $\Delta T$ is the temperature jump across the relevant length $H$, $\kappa$ is the diffusivity of temperature, and $\eta$ is the viscosity. For planetary ice shells, $\eta$ is commonly taken to be $\eta_b$, the basal viscosity of the shell. To complicate matters, the effective ice viscosity $\eta$ is not well prescribed and varies with the material temperature, the magnitude of the shear stress applied to the ice, and the size of the ice grains \citep{Goldsby+Kohlstedt2001, durham2001}. This means that the history and grain size of Europa’s ice shell may also affect its future evolution. 

How did this coupled ice-ocean system evolve in the first place? Either the ice shell must have melted from below, or frozen from the top down. In its early stages, Europa would likely have been a mixture of ice and rock, with an ice shell eventually forming from the top down at Europa's surface \citep{lewis1971,bierson2020}. Further, \citet{Cassen+1979} and  \citet{cassen1980} suggest that if the initial state is an entirely frozen shell, an ocean would be unlikely to form. Thus, for this study, we consider the second, top-down freezing scenario.

Several past studies have considered the steady-state thickness of Europa’s ice shell and the mechanisms of heat transport across it \citep[e.g.,][]{ojakangas1989, mckinnon1999, hussmann2002, tobie2003, moore2006, howell2021}. Fewer studies have considered a growing shell, represented by a moving boundary at the ice-ocean interface. Some of these have considered either a purely conductive \citep{quick2015} or convective \citep{green2021} shell, without considering the transition in heat transport from conduction to convection as the shell thickens.

Here, we explore the temporal evolution of an ice shell with a moving lower boundary and an evolving surface temperature, and we identify the temporal transition between conductive and convective regimes. The transition between a conductive and convective shell has previously been considered in different setups -  with a fixed domain thickness \citep{mitri2005}, without the effect of latent heat \citep{peddinti2019}, or under a  convective parametrization \citep[][who consider how Europa's orbit influences its shell thickness]{hussmann2004}. Further, recent studies of the growth of ice shells on Europa \citep[e.g.,][]{buffo2021, buffo2021b} and Triton \citep{hammond2018} have considered top-down growth in the presence of solutes, with fixed surface temperature boundary conditions in the absence of solid-state convection. Since the mode of heat transport carried across the shell is coupled to its thickness, understanding the transition between conductive and convective transport may be key to understanding the final state of the system, as well as to the origin of the differing Europan surface morphologies. Since this transition is affected by the rheology of the ice, we explore these nuances and analyze how they may affect the ice-shell thickness. In the next section, we discuss the role of viscosity in a solid-state system. In Section 3, we describe the setup of our one-dimensional model. In Section 4, we discuss our results. Finally, we conclude by presenting a novel method to estimate magnitudes of vertical ocean velocities from ice thickness measurements and describe how our results may be used in context with future radar observations from the upcoming Europa \textit{Clipper} mission.

\section{Rheology and convection}\label{sec:viscosity}
It is well known from observations of glaciers and ice sheets, and from laboratory experiments, that ice deforms plastically under a persistent shear stress.
The relationship between deviatoric (i.e., trace-free) stress ($\mb{\sigma}$) and strain rate($\mb{\dot\varepsilon}$) is generally not linear, as it would be in a Newtonian fluid.
Yet several past works that have explored the possibility of convection in Europa's ice shell have treated the ice as though it were an isotropic Newtonian fluid with some assumed (possibly temperature-dependent) viscosity \citep[e.g.,][]{mckinnon1999, tobie2003}. Several other studies considering icy satellites have considered the effects of more complex rheologies \citep[e.g.,][]{barr2004, barr2005, barr2007}.

\citet{Goldsby+Kohlstedt2001} frame their classic synthesis of previous experimental work via relations of the form 
\begin{equation}\label{eq:ssrel}
\dot\varepsilon = A\sigma^n d^{-p}\exp\left(-\frac{Q+PV}{RT}\right),
\end{equation}
in which $\sigma$ is some scalar invariant of the stress tensor $\mb{\sigma}$, $\dot\varepsilon$ is a corresponding invariant of $\mb{\dot\varepsilon}$, $d$ is a characteristic grain size (presuming polycrystalline ice), $T$ is the ice temperature, $P$ its pressure, $Q$ and $V$ are activation energy and volume for creep, $R$ is the gas constant, and $A$ is a coefficient independent of these other quantities.
It is found that in most regimes that can be studied experimentally, the exponent $n$ is greater than unity, making the relation \eqref{eq:ssrel} between stress and strain rate nonlinear.
But no single value of $n$ describes all of the data that \citet{Goldsby+Kohlstedt2001} review; rather, $n$ is approximately constant only within restricted ranges of $\sigma$, with a tendency for larger $n$ in higher stress regimes.

Although difficult to test experimentally, it is believed on theoretical grounds that $n=1$ (implying a linear rheology) at very low stresses and strain rates, where the dominant mechanism becomes diffusion creep.
This mechanism therefore provides a lower bound to the rate of plastic flow at a given stress, and therefore an upper bound to the effective viscosity.
As formulated by \cite{Herring1950}, diffusion creep involves migration of point defects that exist for irreducible thermodynamic reasons.
But because it is a diffusive process, the rate of migration depends on grain size, leading to the factor $d^{-p}$ in eq.~\eqref{eq:ssrel}.
Since $n=1$, the reciprocal of the coefficient of $\sigma$ on the right side of eq.~\eqref{eq:ssrel} has the dimensions of viscosity, and the relation is easily promoted to one between tensors, $\mb{\sigma}=\eta\mb{\dot\varepsilon}$ with effective viscosity $\eta\propto d^p$. 
From eq.~(4) of \citet{Goldsby+Kohlstedt2001}, but neglecting the boundary-diffusion term, which is unimportant for grains larger than about a millimeter at $T>100\unit{K}$, 
\begin{equation}\label{eq:visc1}
    \eta = \left[\frac{42 V_m}{RT d^2}D_{0,V}\exp(-Q_V/RT)\right]^{-1}\\
\end{equation}
\begin{equation*}
    \qquad\qquad\qquad\approx \begin{pmatrix} 
    7\times 10^{14} & @\ T= 273\unit{K} \\
    2\times 10^{15} & @\ T= 263\unit{K}
    \\
    8\times 10^{19} & @\ T= 187\unit{K} \\
    1\times 10^{34} & @\ T= 100\unit{K}
    \end{pmatrix} \times\left(\frac{d}{1\unit{mm}}\right)^2\unit{Pa\ s}.\nonumber
\end{equation*}
For the numerical viscosities on the second line, we have used the values of the parameters from \citet{Goldsby+Kohlstedt2001}'s Table~6:
$V_m=1.97\times10^{-5}\unit{m^3}$, $D_{0,V}=9.1\times 10^{-4}\unit{m^2\,s^{-1}}$, $Q_V=59.4\unit{kJ\,mol^{-1}}$.
The temperatures shown are approximately the melting point, 263~K, the mean temperature of the surface in equilibrium with sunlight, and the average of the melting temperature and surface temperature. Note that other processes may be at play particularly at temperatures close to the melting point, described by \citet{Goldsby+Kohlstedt2001} with a different activation energy, probably reducing the estimated effective viscosity.

Both eqs.~\eqref{eq:ssrel} \& \eqref{eq:visc1} depend explicitly on the grain size.
It seems likely that in an ice shell that has been worked by tides and perhaps convection for millions or billions of years, the grain size should have reached an equilibrium between distortion/fracture and recrystallization.
\cite{jacka1994steady} have fit the following the empirical relation to observations of grain size in glaciers and ice sheets at temperatures $\ge -10^\circ\unit{C}$:
\begin{subequations}\label{eqs:sizes}
\begin{equation}\label{eq:grainsize}
d^2 \approx  0.07\unit{mm^2}\left(\frac{\sigma}{\mathrm{MPa}}\right)^{-3}
\end{equation}
If one fits only the laboratory data that \cite{jacka1994steady} tabulate for $-5^\circ\unit{C}$ and $-10^\circ\unit{C}$, one gets a shallower dependence:
\begin{equation}\label{eq:grainsize1}
    d^2\approx 0.35\unit{mm^2}\left(\frac{\sigma}{\mathrm{MPa}}\right)^{-1.25}
\end{equation}
\end{subequations} 

On the assumption that the ice sheet is thinner than the underlying ocean, the tidal strain in the sheet should be $\approx 10^{-5}$ (the rise and fall of the tide alters the local radius of curvature of the sheet, alternately stretching and compressing it); using the elastic and bulk moduli, this corresponds to an r.m.s. differential stress  $\approx 1.2\unit{bar}$ (i.e., $0.12\unit{MPa}$). Putting the latter into eq.~\eqref{eq:grainsize} predicts $d\approx 6.4\unit{mm}$, while the shallower relation would give $d\approx 2.2\unit{mm}$. We base these predicted grain sizes on the tidal stresses rather than the stresses associated with convection (if it occurs) because we expect the latter are likely to be smaller \citep[as in][]{mckinnon1999}.   

The time to reach the equilibrium grain size can be estimated from the rate at which grains grow in the absence of stress, which \cite{jacka1994steady} also present. These rates are sensitive to temperature, ranging from $0.4\unit{mm^2\,yr^{-1}}$ at $-3^\circ\unit{C}$ to $4\times10^{-4}\unit{mm^2\,yr^{-1}}$ at $-60^\circ\unit{C}$. As an upper bound on the grain-equilibration timescale for the parts of Europa's ice sheet that might convect is therefore $(6.4\unit{mm})^2/(4\times10^{-4}\unit{mm^2 yr^{-1}})\approx0.1\unit{Myr}$. Since this is much less than the age of the moon or even Europa's current surface \citep{pappalardo1999}, it seems likely that the grain size should be in equilibrium. We note that \citet{barr2007}, using a more advanced setup, find equilibrium grain sizes of O(1)~mm to O(100)~mm, which are not inconsistent with our findings. With eq. \eqref{eq:visc1}, this would predict that the viscosity should be in the range $10^{15}$ to $10^{17}\unit{Pa\mhyphen s}$ at temperatures above $263\unit{K}$, depending somewhat on which of the grain-size prescriptions~\eqref{eqs:sizes} one uses. 

These values, however, probably overestimate the effective viscosity because they are based on a single physical mechanism (grain-boundary diffusion), one that dominates the creep rate only at very low stresses. Other deformation mechanisms may be relevant at the moderate deviatoric stresses characteristic of terrestrial ice sheets and glaciers, and probably also Europa's ice shell. These mechanisms are described by exponents $n>1$ in eq.~\eqref{eq:ssrel}, so that the strain rate varies nonlinearly with the stress, and the effective viscosity is considered to be $\eta_{\rm eff}\equiv\sigma/\dot\varepsilon$.
In particular, field data collected from glaciers and ice sheets are generally well described by Glen's law $\dot\varepsilon\propto\sigma^3$ \citep{glen1955, weertman1983creep}. To fix the constants, we adopt the form inferred by \cite{weertman1983creep}:
\begin{equation}\label{eq:Glen}
    \eta_{\rm eff} \approx 10^{15}\left(\frac{\sigma}{1\unit{bar}}\right)^{-2}\exp\left[\frac{Q}{R}\left(\frac{1}{T}-\frac{1}{263\unit{K}}\right)\right]
\end{equation}
with activation energy $Q=137\unit{kJ\,mol^{-1}}$ for $T>263\unit{K}$, and otherwise $Q=60\unit{kJ\,mol^{-1}}$.
This empirical law makes no reference to grain size.
At stresses $\sim 1\unit{bar}$ and temperatures above $263\unit{K}$, eq.~\eqref{eq:Glen} predicts $\eta_{\rm eff}\lesssim 10^{15}\unit{Pa\mhyphen s}$, in accord with the data summarized in Fig.~3 and Table~1 of \cite{weertman1983creep}. This means that the effective viscosity can be related to the stress applied to the ice shell. 

Since Europa's ice shell experiences a steep temperature gradient across its thickness, the temperature dependence of viscosity affects the critical Rayleigh number at which the shell may convect \citep[i.e.,][]{stengel1982, solomatov1995}. Past work has taken this critical Rayleigh number to be $Ra_c = 10^6$ \citep{mckinnon1999}. Further, more recent work has shown how the critical Rayleigh varies depending on both internal heating or lack thereof, and with or without a temperature-dependent viscosity \citep{jain2022}. For isoviscous convection, internal heating does not significantly affect the value of the critical Rayleigh number. However, when internal heating is included along with 
a temperature-dependent viscosity, the critical Rayleigh number may be lower in the internally-heated case \citep{jain2022}, although for the values used here, we find from our own linear-stability calculations based on eq.~\eqref{eq:Glen} that $Ra_c = O(10^6)$ suffices. Note that if $Q = 137$~kJ mol$^{-1}$ (for a reference temperature tied to the freezing point of water), the critical Rayleigh number may be an order of magnitude larger. We consider these influences of temperature-dependent viscosity on the critical Rayleigh number in our model described next. 

\section{One-dimensional Thermodynamic Model}\label{method}

To consider the temporal evolution of the shell, we assume a pure ice shell grown from the surface downwards, such as would happen if ice and rock differentiated, and then the ice melted and refroze. This allows one to study the transition from a conductive to a convective shell without having to invoke parametrizations for convection, as would be required for an ice shell melting at its base. 

The governing equations are based on \citet{maykut1971}, who developed a one-dimensional thermodynamic model for sea ice growth in the Arctic Ocean. We note that these equations apply \textit{only} in a conductive regime. Heat transfer across the shell is governed by the heat equation for conductive transfer with an internal heat source:
\begin{equation}\label{eq:heat}
    \rho c_p\frac{\partial T}{\partial t}=\frac{\partial}{\partial z}(k\frac{\partial T}{\partial z})+I,
\end{equation}
where $z$ is depth below the surface, $\rho$ is the density of ice, $c_p$ is the specific heat capacity of ice, $k$ is the thermal conductivity, $T$ is temperature, and $I$ is a specified rate of internal heating. 

For planetary ice shells, the internal heating due to tides is commonly prescribed as a function of viscosity, itself a function of depth \citep[e.g.,][]{ojakangas1989}. However, we generally choose to consider a constant internal heating rate throughout given the uncertainties in ice rheology.  We do describe simulations of a depth-dependent internal heating rate in particular contexts, utilizing the following prescription, as in \citet{ojakangas1989}:
\begin{equation}
    I=\frac{2\mu\dot{\epsilon}^2}{\omega}[\frac{\omega\eta/\mu}{1+(\omega\eta/\mu)^2}],\label{I_var}
\end{equation} where $\mu=4\times10^9$~Pa is the ice rigidity, $\dot{\epsilon}=2\times10^{-10}$~s$^{-1}$ is the strain rate, $\omega=2.048\times10^{-5}$~s$^{-1}$ is the forcing frequency (tidal), and $\eta$ is the ice viscosity. Here, the viscosity is commonly taken to be:
\begin{equation}
    \eta = \eta_b exp[\frac{Q}{R}(\frac{1}{T}-\frac{1}{T_0})]\label{eta_var},
\end{equation} where $Q = 59000$~J mol$^{-1}$, $R$ is the gas constant, and $T_0$ is the melting temperature \citep{ojakangas1989, nimmo2007}. We take $T_0$ and $\eta_b$ to be the temperature and viscosity at 263~K, to avoid a jump in activation energy at higher temperatures.

We note that the density, specific heat capacity, and thermal conductivity are all temperature-dependent \citep[see][for a nice description]{Carnahan+2021}, though many studies of Europa's ice shell consider only fixed values \citep[e.g.,][]{hussmann2002, quick2015}, as we do here. Throughout, we take the density of ice $\rho_i = 920$~kgm$^{-3}$, the specific heat capacity of ice as $c_p=2100$~Jkg$^{-1}$K$^{-1}$, and the ice conductivity as $k=2.2$~Wm$^{-1}$K$^{-1}$ (approximate values for $T=273$~K). We note that steady-state shell thicknesses calculated from a constant conductivity are reduced by at most $\sim$40\% from what they would be if calculated with a temperature-dependent conductivity, where $k =612/T$~Wm$^{-1}$K$^{-1}$, as in \citet{Carnahan+2021}. 

\begin{figure}   \includegraphics[width=0.6\textwidth]{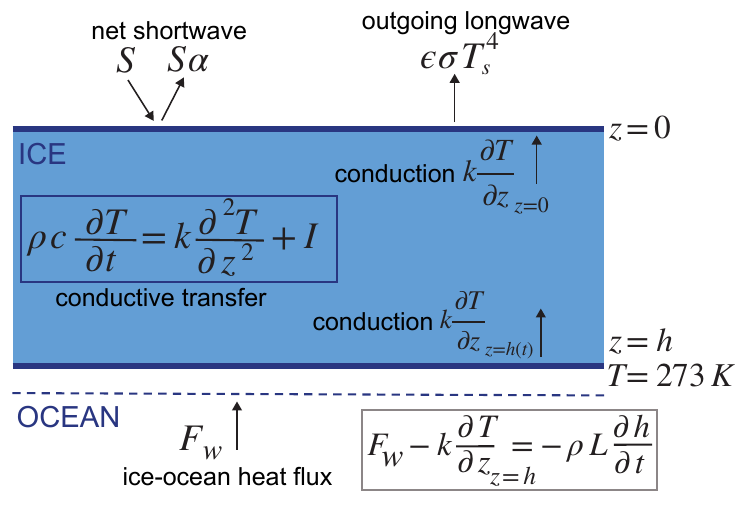}
        \caption{Schematic of the 1-D model for Europa's ice shell growth. Heat is transferred across the shell via molecular conduction, with a potential internal source term $I$. At the upper boundary there is a balance between net shortwave radiation (total incoming shortwave radiation $S$ and albedo $\alpha$), conductive heat transfer into the upper boundary, and outgoing longwave radiation ($\epsilon$ is emissivity, $\sigma$ is the Stefan-Boltzmann constant, and $T_s$ is the temperature at $z=0$, the upper surface). The bottom boundary, $z=h(t)$, is tied to the freezing temperature. An ice-ocean heat flux $F_w$ is applied at the base of the shell. The shell grows according the grey-boxed equation, $F_w - k\frac{\partial T}{\partial z}_{z=h}=-\rho L \frac{\partial h}{\partial t}$, where $k$ is the thermal conductivity, $\rho$ is the ice density, $L$ is the latent heat of fusion, $h$ is the ice thickness, and $t$ is time.}\label{setup}
\end{figure}

The boundary condition at the lower surface is dictated by the melting temperature of water:
\begin{equation}\label{eq:lower}
    T(z=h)=273~K,
\end{equation}
where $h$ is the depth of the ice shell. We neglect here the pressure and salinity dependence of the melting temperature; these will vary by only a few degrees for the range of pressures and salinities appropriate to the Europan case.
The boundary condition at the upper surface is given by the radiative balance:
\begin{equation}\label{eq:upper} F_r(1-\alpha_i)+k\frac{\partial T}{\partial z}\rvert_{z=0}-\epsilon\sigma T_s^4=0,
\end{equation}
where $F_r=10$~Wm$^{-2}$ is an approximate average value for the incoming shortwave solar radiation \citep{ashkenazy2019}, $\alpha_i\approx0.6$ is an approximate value for ice albedo \citep[as in][though the albedo may be higher and is spatially-varying]{buratti1983}, $\epsilon=0.94$ is emissivity \citep{ashkenazy2019}, $\sigma=5.67\times10^{-8}$~Wm$^{-2}$K$^{-4}$ is the Stefan-Boltzmann constant, and $T_s$ is the temperature at the upper surface ($z=0$). The first term describes the amount of solar radiation absorbed by the surface, the second term describes the conductive heat transfer to the surface, and the third term is the outgoing longwave radiative flux (Figure \ref{setup}).  

The bottom boundary grows according to the Stefan condition:
\begin{equation}\label{eq:stefan}
    F_w - k\frac{\partial T}{\partial z}\rvert_{z=h}=-\rho L\frac{\partial h}{\partial t},
\end{equation}
where $F_w$ is the incoming heat flux from the ocean and $L=3.3\times10^5$~Jkg$^{-1}$ is the latent heat of fusion. This equation relates the growth of the ice shell to the incoming ocean heat flux and the conductive flux out of the lower boundary (Figure \ref{setup}). 

The initial ice shell thickness $h_1$ is calculated from a heat balance:
\begin{equation}
    F_r(1-\alpha_o) - \epsilon\sigma T_s^4=-\rho L\frac{\partial h}{\partial t}.
\end{equation}
$\alpha_o\sim0.07$ is the ocean albedo \citep[e.g.,][]{perovich2012}, $T_s=273$~K, and $t$ is integrated from $t=0$ to $t_1=10^6$~seconds. $F_w$ is neglected here because it is negligible compared to the other terms prior to sea ice formation.

Then, the initial temperature profile condition is taken to be: 
\begin{equation}
     T(z,t_1)= 273 + (T_s-273)(1-z/h_1),
\end{equation}
where $T_s$ is now calculated from the surface energy balance, assuming a linear profile.

The conductive equations apply only until the point at which convective transfer becomes the dominant mode of heat transport. Thus, it is necessary to determine at each time step whether the shell may be prone to convective instability. This is inferred from the Rayleigh number, $Ra\equiv\frac{\rho g\beta\Delta TH^3}{\eta_b\kappa}$, where now $\Delta T$ describes the temperature jump across the ice shell, $H$ is the thickness of the ice shell, and $\eta_b$ is viscosity of the shell at T=263~K (where the activation energy is $\sim$59 kJ mol$^{-1}$). Here, $\beta=1.7\times10^{-4}$~K$^{-1}$ is the coefficient of thermal expansion \citep[][taken at 273~K]{kirk1987}, and $g=1.3$~ms$^{-2}$ is Europa's gravitational constant. 

The transition from conductive to convective heat transfer occurs at a critical Rayleigh number. For the Europan case with a temperature-dependent viscosity, this has been taken to be $Ra_c=O(10^6)$ \cite[e.g.,][]{solomatov1995, mckinnon1999}, though it has been argued that internal heating may reduce the critical Rayleigh number \citep{jain2022}. For an Arrhenius dependence of viscosity on temperature as in eq. (5) (with Q = 60 kJ/mol), and a uniform internal heating rate, we find via a linear stability analysis using Dedalus \citep{Burns+2020} that $Ra_{c}$ decreases by roughly one order of magnitude as the importance of internal heating increases: from $Ra_{c} = O(10^6)$ for $I=0$ to $O(10^5)$ for $I=2k\Delta T/h^2$.  For a constant conductivity $k$, as we have assumed, the conductive temperature profile is monotonic in $z$ only when $Ih^2\le 2k\Delta T$, so the latter value of $I$ is an upper bound to the volumetric heating at a given shell thickness $h$. Then, when $I$ has approximately half of this maximum value, $Ra_{c}\sim O(10^6)$, the value that we take in this analysis. In a model where the internal heating rate is not uniform but rather strongly concentrated toward the base---following eq.~\eqref{I_var}---we expect that $Ra_{c}$ should have an intermediate value, because the temperature profile above the strongly heated region resembles that of a model without internal heating but with an increased basal flux.

\subsection{Numerical Method}

To solve equations (\ref{eq:heat})-(\ref{eq:stefan}), we use a centered-difference scheme in space and forward-difference scheme in time with an adaptive time step. There are 101 vertical grid points, and the coordinate system is rescaled to $(\hat{T}=T, \hat{z}=z/h)$ to account for the moving boundary. The rescaled equations then become:

\begin{equation}
    \hat{T}(\hat{z},0)= 273 + (\hat{T_s}-273)(1-\hat{z}), 
    \textrm{for the initial condition}.
\end{equation}
Note that $\hat{T_s}=T_s$ since the upper boundary does not move with time. The boundary conditions are:
\begin{equation}
    \hat{T}(\hat{z}=1)=273~K
\end{equation}
and
\begin{equation}
    F_r(1-\alpha_i)+\frac{k}{h}\frac{\partial\hat{T}}{\partial\hat{z}}\rvert_{z=0} - \epsilon\sigma \hat{T}^4(\hat{z}=0) = 0.
\end{equation}
The heat equation is:
\begin{equation}
    \frac{\partial\hat{T}}{\partial t}-\frac{\dot h\hat{z}}{h}\frac{\partial\hat{T}}{\partial\hat{z}}=\frac{\kappa}{h^2}\frac{\partial ^2 \hat{T}}{\partial \hat{z}^2}+\frac{I}{\rho c_p},
\end{equation}
where $\dot{h}\equiv dh/dt$, and the Stefan condition is:
\begin{equation}
    F_w-\frac{k}{h}\frac{\partial\hat{T}}{\partial\hat{z}}\rvert_{\hat{z}=1}=-\rho L \frac{d h}{d t}.
\end{equation}
The Rayleigh number is computed at each time step. When the critical Rayleigh number hits the critical value, the shell is assumed to enter a convective regime, and our model no longer applies. For some parameters, the shell may evolve to a conductive steady-state thickness, never reaching the critical Rayleigh number. 

\subsection{Parameters}
There are several free parameters in the system, specifically the ice-ocean heat flux $F_w$, the reference ice shell viscosity $\eta_b$, and the rate of internal heating $I$. We vary these across a range of values relevant to the Europan system and discuss the effects on the ice shell thickness. The ice-ocean heat flux $F_w$ (resulting from radiogenic decay or tidal heating from the core) varies from $O(0.01)-O(0.1)$~Wm$^{-2}$ \citep[e.g.,][]{thompson2001, hussmann2002}. The viscosity value $\eta_b$ varies from $10^{13}-10^{17}$~Pa s, in line with other studies \citep[e.g.,][]{ashkenazy2018, howell2021}. Finally, we take the range of internal heating in the ice shell $I$ to be between $10^{-5}$ and $10^{-7}$~Wm$^{-3}$, as in \citet{showman2004}; realistically, the rate of internal heating will likely depend on the viscosity \citep[also shown in e.g.,][]{showman2004} and described above. Varying these parameters gives an indication of where we can expect Europa's ice shell in a conductive or convective regime.  

\section{Results}\label{results}
\subsection{Steady-State Analytical Solution}

In a steady-state, the equations can be solved analytically.  Then equation \ref{eq:heat} with constant conductivity becomes:
\begin{equation}
    k\frac{\partial ^2 T}{\partial z^2}=-I.
\end{equation}
For a fixed surface temperature ($T_s$), the steady-state solution is then:
\begin{equation}
    T(z) = T_s + (273 - T_s)\frac{z}{h}+\frac{I}{2k}z(h-z).
\end{equation}
The Stefan condition (Equation \ref{eq:stefan}) becomes:
\begin{equation}\label{eq:stef_steady}
    F_w  = k(\frac{273 - T_s}{h}-\frac{Ih}{2k}),
\end{equation}
and the energy balance at the upper surface (Equation \ref{eq:upper}) becomes:
\begin{equation}\label{eq:surface_steady}
    F_r(1-\alpha_i)+k(\frac{273 - T_s}{h}+\frac{Ih}{2k}) - \epsilon\sigma T_s^4 = 0,
\end{equation}
where now Equations \ref{eq:stef_steady} and \ref{eq:surface_steady} can be solved simultaneous for $h$ and $T_s$. We stress that these solutions will only be valid provided that the system is in a conductive regime; this is why calculating the Rayleigh number at each time step is necessary. For systems which do remain in the conductive regime, the steady-state analytical solution provides a means by which to validate the numerical results. 
\begin{figure}
\includegraphics[width=\textwidth]{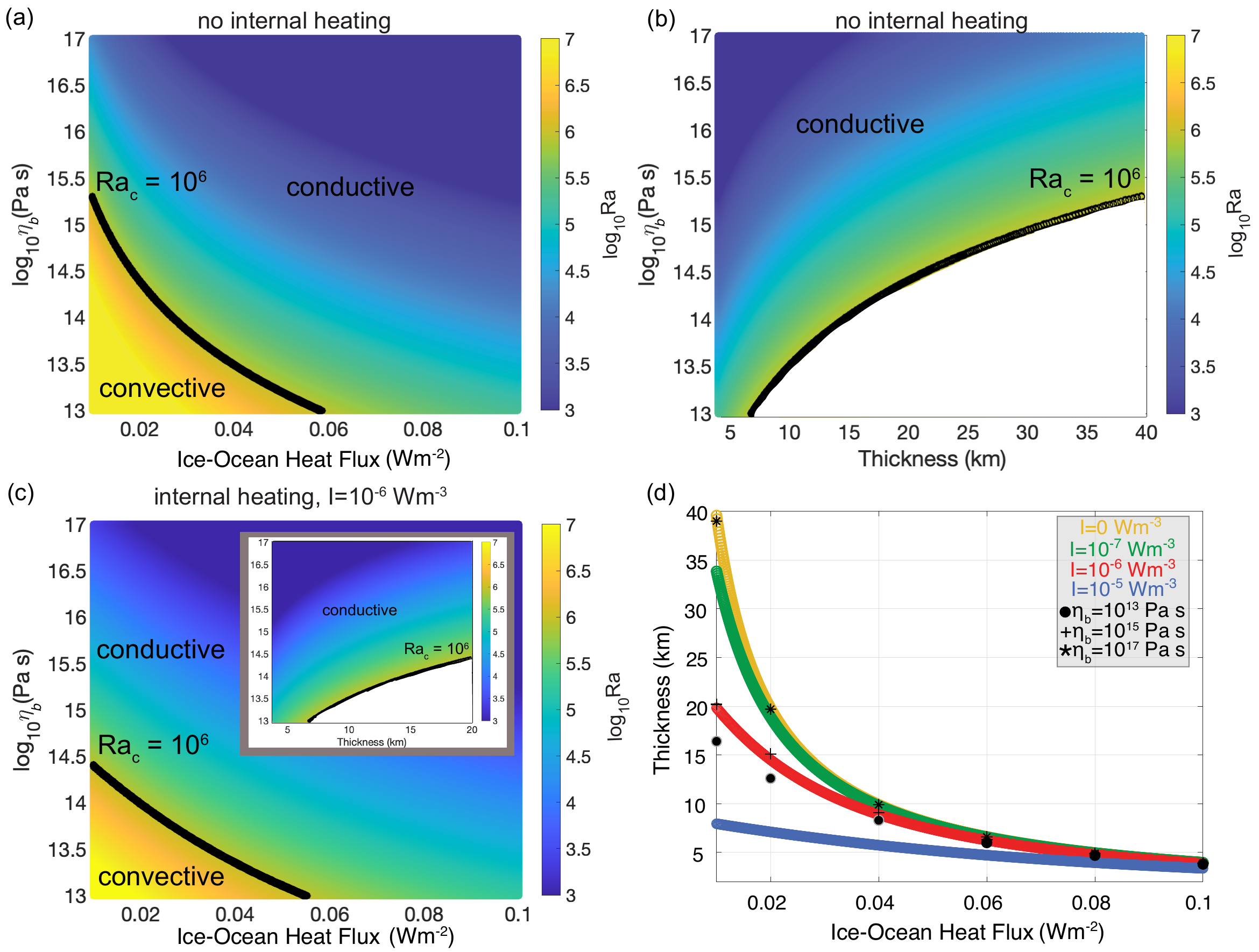}
        \caption{Values of (a) ice-ocean heat flux and (b) thickness under which a conductive steady state is expected at particular basal viscosities in an ice shell with no internal heat source. The Rayleigh number is shown in color; here the critical Rayleigh number is taken to be 10$^6$. The white section of the color plot in (b) as well as in (c) corresponds to the region for which steady-state conductive thicknesses would exceed the critical Rayleigh number. (c) Values of ice-ocean heat flux and thickness (inset) under which a conductive steady state is expected at particular basal viscosities in an internally-heating ice shell ($I=10^{-6}$~Wm$^{-3}$). Rayleigh number is in color. Related steady-state conductive thicknesses are shown in the inset.(d) Relationship between ice shell thickness and ice-ocean heat flux for different values of internal heating (color) in steady-state. Depth-dependent internal heating following the prescription in Equation \ref{I_var} for particular reference viscosities are shown by the black dots ($\eta_b=10^{13}$ Pa s), plus signs ($\eta_b=10^{15}$ Pa s), and stars ($\eta_b = 10^{17}$~Pa s). The black dots with the gray outline indicate ice thicknesses which would be unstable to convection for the given reference viscosity and a critical Rayleigh number of O(10$^6$).}. \label{fig3}
\end{figure}

In a steady state with no internal heating, whether the ice shell remains in a conductive state is dependent on the ice-ocean heat flux and the ice rheology (Figure \ref{fig3}a, b). Similar ideas have also been explored by \citet{mckinnon1999} and \citet{mitri2005}, exemplifying the necessity to constrain $\eta_b$ and grain size for the stresses relevant to Europa's ice shell. Including internal heating increases the range of values over which a shell can be conductive. This is because the volumetric internal heating term leads to a thinner ice shell, which will then take longer to convect for the same critical Rayleigh number ($Ra_c = 10^6$). For example, for a constant value of internal heating ($I = 10^{-6}$~Wm$^{-3}$), for an ice-ocean heat flux $F_w\gtrsim$0.06~Wm$^{-2}$ and basal viscosity $\eta_b\gtrsim$10$^{14.5}$~Pa s, the shell will always be in a conductive state (Fig. \ref{fig3}c), compared to a case with no internal heating where the shell will remain conductive for a similar value of ice-ocean heat flux but values of basal viscosities $\eta_b\gtrsim$10$^{15.5}$~Pa s (Fig. \ref{fig3}a). Further, since the rate of internal heating is a volumetric term, the thicker the ice shell becomes, the larger is the effect that the heating has (Figure \ref{fig3}d). For example, subject to an internal heating rate of $I=10^{-6}$~Wm$^{-3}$ and ice-ocean heat flux of $F_w=5\times10^{-2}$, the shell will reach 7360~m thick compared to 7900~m thick without internal heating, where as for a shell subject to an ice-ocean heat flux of $F_w=10^{-2}$~Wm$^{-2}$, the same rate of internal heating reduces the shell thickness to about 20,000~m compared to about 40,000~m without internal heating. Further, at low ice-ocean heat fluxes in a conductive steady state, an ice shell subject to an internal heating rate of $I=10^{-6}$~Wm$^{-3}$ leads to an ice shell about two times thinner than in the absence of internal heating (Fig. \ref{fig3}d), provided the shell remains conductive.

Ice shells subject to depth-dependent internal heating (Equation \ref{I_var}) result in similar steady-state thicknesses (provided the shells remain conductive) to those subject to constant internal heating. For example, steady-state ice shell thicknesses for reference viscosities of 10$^{13}$~Pa s and 10$^{15}$ Pa s and depth-dependent internal heating closely align with those attained for a constant heating rate of $I=10^{-6}$~Wm$^{-3}$ (Figure \ref{fig3}d). This indicates that using a constant internal heating rate of $\sim I=10^{-6}$~Wm$^{-3}$ is a reasonable approximation to make when considering the effect of tidal heating on the Europan ice shell. In the case of large-enough viscosity (i.e., $\eta_b=10^{17}$~Pa s) the effect of a depth-dependent internal heating will be basically negligible (Figure \ref{fig3}d). 

Further, for a conductive case with no internal heating, each ice-ocean heat flux corresponds to a particular steady-state shell thickness, which may be useful for inferring parameters relevant to ocean dynamics. For example, shown by the above steady-state solutions, the equilibrium thickness for a conductive shell in thermodynamic balance subject to a ice-ocean flux of 10$^{-2}$~Wm$^{-2}$ is 10 times larger (about 40~km) than that subject to an ice-ocean heat flux of 10$^{-1}$~Wm$^{-2}$ (about 4~km). Even in the case of depth-dependent internal heating, each steady-steady ice thickness can still be related to an ice-ocean heat flux, provided that the reference viscosity is known (Figure \ref{fig3}d) and that the shell is in a conductive state.

These findings highlight the importance of understanding how an ice-shell responds to a heat flux from the ocean. Spatial variations in thickness across the shell may be due to differences in surface temperature (ultimately due to the incoming shortwave radiation from the sun) \citep[e.g.,][]{ashkenazy2018, ashkenazy2019}, but are also controlled by the incoming heat flux from the ocean. We further note that this flux has previously been treated as a constant in studies of steady-state thickness of Europa's ocean \citep[e.g.,][]{ojakangas1989}, but whether the ice-ocean heat flux is spatially variable, and how the ocean may dissipate energy generated by tidal heating in the core, is not well constrained. For example, if Europa's purported geyser activity can be related to sporadic hydrothermal venting on Europa's seafloor and subsequent ocean overturns \citep{shibley2021}, this would indicate an almost certainly spatially-variable ice-ocean heat flux.

This means that if one had a particular observation of shell thickness (such as from \textit{Clipper}) and knew the ice viscosity, one could infer the vertical ice-ocean heat flux reaching the shell base, on the assumption that the thickness had reached a steady state. This may have implications for inferring possible ocean velocities, as discussed in the following section.

\subsection{Surface Temperature}

The growth of Europa's ice shell is driven by the rate at which heat is conducted away from the bottom of the shell compared to the rate at which heat from the ocean is transported into the shell (Eq. \ref{eq:stefan}), while the surface temperature results from a balance between the net shortwave radiation and heat transported via conduction into the surface of the shell and the outgoing longwave radiation (Eq. \ref{eq:upper}). Regardless of ice-ocean heat flux or ice rheology, Europa's surface temperature rapidly approaches an equilibrium value of 93-94K (e.g., after 1 million years, the surface temperature has nearly reached its final value, Figure \ref{fig4}a), although the exact surface temperature depends on the ice-ocean heat flux as well as the prescribed value of shortwave radiation and internal heating. The steady-state surface temperature varies from 93.8~K for $F_w = 10^{-1}$~Wm$^{-2}$ and $I=10^{-5}$~Wm$^{-3}$ to 93.1~K when $F_w =10^{-2}$~Wm$^{-2}$ and $I=0$~Wm$^{-3}$, and depends on the ice-ocean heat flux and rate of internal heating. Higher surface temperatures correspond to larger ice-ocean heat fluxes for the same rate of internal heating, and vice versa, since the outgoing longwave radiation must balance the incoming shortwave radiation and conductive heat flux across the shell. Here, our selection of the value for the incoming shortwave radiation $S=10$~Wm$^{-2}$ sets the value of surface temperature we obtain.

Further, it can be shown using the heat capacity and conductivity of the ice that the diurnal variation in the surface temperature should be less than $\pm 1\unit{K}$ once the ice sheet is thicker than a few tens of meters; and furthermore that the time-averaged surface temperature differs even less than this from what one gets by replacing $T_s$ with its time average in the surface boundary condition, as we do here. For a more complete discussion of global variations in Europa's steady-state surface temperature, see \citet{ashkenazy2019}.

In general, Europa's surface temperature reaches a steady-state value much more rapidly than the ice thickness reaches a steady-state value. Since the surface temperature equilibrates quickly compared to the age of the ice shell, this lends credence to the approach of many studies which simply prescribe a constant-surface-temperature boundary condition \citep[e.g.,][]{mitri2005, quick2015, buffo2021}, rather than solving for the full radiative balance. However, for ice shells subject to large ice-ocean heat fluxes (e.g., $F_w=10^{-1}$~Wm$^{-2}$, see Figure \ref{fig4}b), which approach a steady-state thickness in a similar order of time as the surface temperature approaches an equilibrium value, it may be necessary to include the full surface mixed boundary condition in the solution.

\subsection{Time to Reach Steady-State}

Along with affecting the dominant mode of heat transport across the ice shell, the ice-ocean heat flux and ice rheology control the time for an ice shell to reach a conductive steady-state thicknesses.

Assuming the basal viscosity is large enough to prevent ice-shell convection, the shells subject to the largest ice-ocean heat fluxes reach a steady-state most quickly (Figure \ref{fig4}b). Shells with lower ice-ocean fluxes take significantly longer to reach equilibrium. For example, a shell subject to an ice-ocean heat flux of $10^{-1}$~Wm$^{-2}$ approaches a conductive steady state (taken here to be where the shell thickness reaches to within 0.1\% of the final thickness), in about 3 million years. For an intermediate value of $F_w = 5\times10^{-2}$~Wm$^{-2}$, the ice shell would approach a conductive steady state in about 13 million years. In fact, a shell subject to an ice-ocean flux of $10^{-2}$~Wm$^{-2}$ would take about 314 million years to reach equilibrium; we note, however, that 314 million years is longer than the inferred surface age of Europa \citep{pappalardo1999}. This means that it may be likely that Europa is subject to an ice-ocean heat flux larger than this. Since the dearth of craters on Europa's surface suggests that its surface has been recycled over Europa's history, it may also be possible that the ice shell has not completed growing. This suggests that a steady-state assumption for the ice shell thickness of Europa \citep[e.g.,][]{ojakangas1989} may not always be a good assumption, and that the temporal evolution of the shell should be considered in studies of shell thickness.

Including internal heating in the ice shell decreases the time for an ice shell to reach a conductive steady state (Figure \ref{fig4}c). This is due to the fact that internal heating reduces the thickness of the ice shell. The temporal effect is more pronounced at low values of ice-ocean heat flux since internal heating causes a more significant decrease in steady-state thickness at these values. For example, for $F_w=10^{-2}$~Wm$^{-2}$ and $I=10^{-6}$~Wm$^{-3}$, the shell approaches an equilibrium thickness in about 48 million years, compared to the 314 million years without internal heating. However, for $F_w=10^{-1}$~Wm$^{-2}$, the shell approaches an equilibrium thickness in about 3 million years, with or without constant internal heating (though it is slightly smaller with internal heating). When both a temperature-dependent conductivity and depth-dependent internal heating are included, the effect becomes more nuanced as a temperature-dependent conductivity acts to thicken the shell, while the internal heating acts to thin it; however, a transition from conduction to convection at $\eta_b = 10^{13}$~Pa s occurs at similar times in all cases shown (Figure \ref{fig4}c).

\subsection{Time to Convective Transition}

Surface observations of cracks indicate that Europa's shell may be thin \citep[e.g.,][]{hoppa1999} and thus conductive, while other diapir-like surface features indicate that the shell may be thick and convecting \citep[e.g.,][]{pappalardo1999}. However, it is possible that a shell may have started out in a thin, conductive state, and then subsequently transitioned to a thicker, convective state \citep[e.g.,][]{mitri2005}. Once in a convective state, a steady-state thickness may be reached by a balance between internal heating and convective heat transport through the shell \citep[e.g.,][]{moore2006, green2021}; if internal heating is commonly assumed to be related to ice viscosity \citep[e.g.,][]{ojakangas1989, moore2006, nimmo2007} via Equation \ref{I_var}, then the rate of internal heating will increase once the shell begins to convect. Here, our results indicate how the ice-ocean heat flux, which controls the ice-shell thickness, and ice rheology govern the time of transition to a convective state.   

\begin{figure}
    \includegraphics[width=\textwidth]{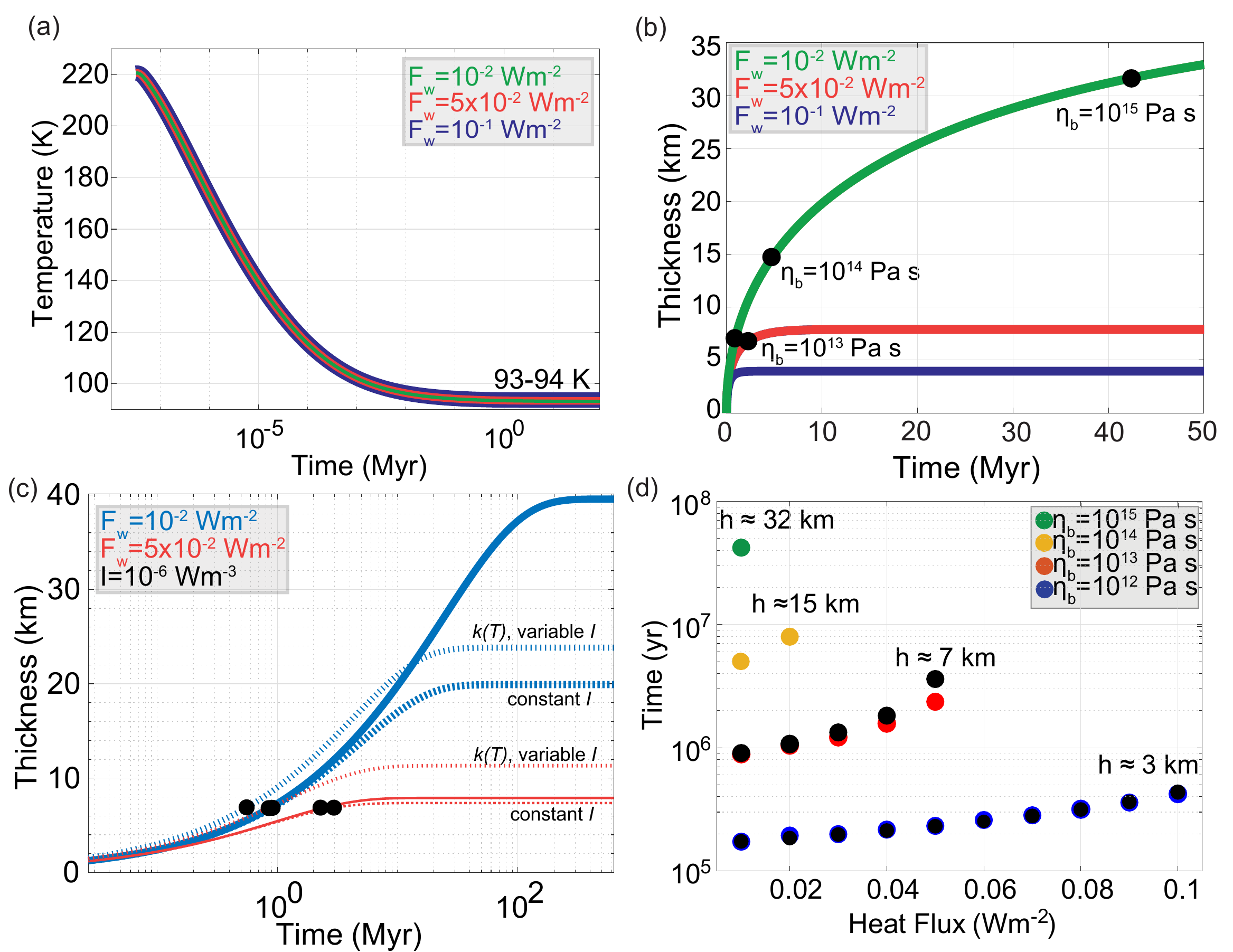}
        \caption{(a) Temporal evolution of ice shell surface temperature for different values of ice-ocean heat flux (colored lines). The surface temperature reaches a fairly constant temperature after about 1 million years. (b) Temporal evolution of ice shell thickness (km) with time (in millions of years). The colored lines correspond to different heat fluxes (green$=10^{-2}$~Wm$^{-2}$, red$=5\times10^{-2}$~Wm$^{-2}$, blue$=10^{-1}$~Wm$^{-2}$). The time at which the shell switches from conductive to convective heat transport (dependent on the basal viscosity, $\eta_b$) is marked by a black dot. (c) Temporal evolution of ice shell thickness (km) with (thick dashed line) and without (solid line) constant internal heating of $I=10^{-6}$~Wm$^{-3}$, subject to an ice-ocean heat flux of $F_w = 5\times10^{-2}$~Wm$^{-2}$ (red) and $F_w = 10^{-2}$~Wm$^{-2}$ (blue). Thin dashed lines correspond to simulations with temperature-dependent conductivity and depth-dependent internal heating using $\eta_b = 10^{13}$~Pa s. The time of transition from conductive to convective heat transport (dependent on the basal viscosity, taken here to be $\eta_b=10^{13}$~Pa s) is marked by a black dot. (d) Initial time at which ice shells subject to different ice-ocean heat fluxes switch between conductive and convective heat transport, for varying basal viscosities (colored dots). Black dots indicate the transition time for shells subject to internal heating (and constant conductivity) at reference viscosities of $\eta_b=10^{12}$~Pa s and $\eta_b=10^{13}$~Pa s. The thickness of the shell at the transition time is labeled.} \label{fig4}
\end{figure}

The ice rheology is a key control on the mode of heat transport, and thus the time and thickness at which an ice shell becomes unstable to convection. For higher values of viscosity, at the same ice-ocean heat flux, the ice shell may grow for a longer time than for lower values of viscosity before transitioning into a convective regime (Figure \ref{fig4}b,d). For example, for $F_w = 10^{-2}$~Wm$^{-2}$, the ice shell will transition to a convective regime after about 42 million years for a basal viscosity $\eta_b = 10^{15}$~Pa s, and in about 1 million years for $\eta_b = 10^{13}$~Pa s (Figure \ref{fig4}b,d). For basal viscosities ranging from 10$^{12}$~Pa s to 10$^{15}$~Pa s in the absence of internal heating, the time at transition ranges from around O(100,000) years to over O(10 million) years, respectively. If a viscosity-dependent internal heating is included, shells subject to lower values of viscosity (i.e., $\eta_b=10^{12}$~Pa s and $\eta_b=10^{13}$~Pa s) will still become unstable to convection at similar times to those in the absence of internal heating. However, for larger values of viscosity (i.e., $\eta_b=10^{14}$~Pa s and $\eta_b = 10^{15}$~Pa s), internal heating will have more of an effect on reducing the shell thickness, thereby keeping the shell in a conductive state (Figure \ref{fig4}d). If the viscosity is extremely large, however, viscosity-dependent internal heating becomes negligible again. Ice shell thicknesses at these transition times range from about 3~km to 32~km, with thinner shells corresponding to lower basal viscosities and thus earlier transitions from a conductive to convective state.  

The ice-ocean heat flux also plays a role in determining the time of the transition given its role it determining the ice thickness. For the same value of basal viscosity, the time at which an ice shell transitions between the conductive and convective states depends on the ice-ocean heat flux. For larger values of $F_w$, the transition between states happens at later times than for smaller values of $F_w$, since the shell will stay thinner for longer when subject to the larger ice-ocean heat flux (Figure \ref{fig4}b,d). However, this effect is small compared to the relative effect of viscosity in controlling the transition time. For example, for $\eta_b = 10^{12}$~Pa s, the transition time ranges from about 200,000 years for $F_w=0.01$~Wm$^{-2}$ to about 400,000 years for $F_w=0.1$~Wm$^{-2}$, whereas a basal viscosity of $\eta_b=10^{13}$~Pa s compared to $\eta_b=10^{12}$~Pa s at a given ice-ocean heat flux increases the conductive-convective transition time by almost an order of magnitude. Thus, if one knew the ice-ocean heat flux and ice rheology, our results would indicate the time and thickness of an ice shell at a transition between a conductive and convective state. This may shed insight on the understanding of Europa's observed surface features. 

Internal heating also delays the onset of convection for $Ra_c = O(10^6)$. For example, for an ice-ocean heat flux of $F_w=5\times 10^{-2}$~Wm$^{-2}$ and a basal viscosity of $\eta_b=10^{13}$~Pa s, the onset of convection will occur at around 2 million years if there is no internal heating, and at around 3 million years if $I=10^{-6}$~Wm$^{-3}$ (Figure \ref{fig4}c). This can also be seen for the viscosity-dependent internal heating results show in in Figure \ref{fig4}d. If shells with internal heating remain in a state of conductive heat transport, they  will ultimately be thinner than shells without internal heating. However, there is some evidence that particular internally-heated shells may begin to convect sooner than those without internal heating \citep{jain2022}, and further work is necessary to determine how differing rates of internal heating and basal heat fluxes can affect the onset of convection. Since internal heating affects the boundary layer structure of a convective ice shell and is likely dependent on the shell viscosity, this internal heating then needs to be taken into account when considering heat transport across the shell and is non-trivial \citep[as in][for the case of a moving boundary]{green2021}. 

Ultimately, the ice-ocean heat flux at the base of the ice shell and ice rheology, along with the possible internal tidal heating, control the temporal growth and steady-state conductive thickness of the ice shell. 

\section{Discussion and Conclusion}\label{discussion}

In order to determine the temporal evolution of a pure ice shell on Europa, we solve the time-dependent equations for conduction across an ice shell, subject to basal heating at the ice-ocean interface, and a radiative balance at the surface. The equilibrium surface temperature depends on the value of ice-ocean heat flux and internal heating; surface temperatures approach their equilibrium values after about 1 million years. We find that the ice rheology and the ice-ocean heat flux control the evolution of the shell in a conductive state in the absence of internal heating. Shells subject to large ice-ocean heat fluxes will remain thinner and are more likely to maintain a conductive state at a given basal viscosity. Shells subject to lower ice-ocean heat fluxes grow thicker and thus are more likely to enter convective states. Shells subject to larger ice-ocean heat fluxes reach a conductive steady state in O(1-10 million) years, while shells subject to lower ice-ocean heat fluxes take O(100 million) years to reach steady state. This raises the question of whether such shells may still be growing at present. Further, the ice-ocean heat flux and ice rheology control the transition to convection, with transition times ranging from O(10$^5$-10$^7$) years at thicknesses of O(1-10)~km, respectively. When  internal tidal heating is included, the shell transitions to a convective state at later times (if at all) than without internal heating, since the shell thickness is reduced. Thus, it is essential to determine how and whether internal heating affects a conductive ice shell, as this may control the transition to convection and subsequent evolution of Europan surface features.

We note also that here we have taken the onset of convective instability in both the presence and absence of internal heating to occur when $Ra>10^6$. However, if the critical Rayleigh number differs, this will affect the range of viscosities and ice-ocean heat fluxes for which the shell remains in a conductive state.  With our setup for $\eta_b \geq 10^{13}$~Pa s for values of $I\leq10^{-6}$~Wm$^{-3}$, if a shell thickness is found to be about 7~km or less, it will always be in a conductive state; thus each value of steady-state shell thickness can be related to an approximate value of ice-ocean heat flux. If a value of shell thickness is found to be larger than $\sim$ 7~km and the value of $\eta_b$ is unknown, it is not possible to use this method to determine $F_w$, since the shell may be in either a conductive or convective state. This emphasizes the necessity for more studies of ice rheology at the strain rates and grain sizes appropriate for Europa's shell, as well as for a method to infer ice rheologies from the surface observations of Europa's ice shell.

\subsection{Implications for the Ocean}

A primary feature of Europa's ice shell is that it is an observable with direct correspondence to the ocean. Since there are no existing or imminently planned in-situ measurements of the ocean, understanding the dynamics and thermodynamics which govern the properties of Europa's ice shell are key to inferring properties from the subsurface.  This approach, combined with the pioneering usage of magnetometer data and inferences of the induced magnetic field to infer properties of Europa's ocean \citep[e.g.,][]{kivelson2000, khurana2002, vance2021}, helps shed light on subsurface ocean dynamics. 

Realistically, if the ice shell is grown from a salty ocean, it will initially evolve as a mushy layer, which is a porous medium consisting of ice and salt water. Much work on the mushy-layer evolution of Arctic sea ice has shown how the salt concentration can influence the thermodynamic growth of the medium \citep[e.g.,][]{wettlaufer1997, feltham2006}. Such work has been extended by \citet{buffo2021b} to the Europan context with a fixed surface temperature boundary condition. However, since the constraints on Europa's salt content are still quite broad \citep[e.g.][]{hand2007, zolotov2001}, we consider here only the case of a pure ice shell, grown from a freshwater ocean.  The presence of salinity effectively increases the heat capacity of the ice and decreases its conductivity \citep{maykut1971}; thus, our results for a fresh shell provide an upper bound on the shell growth rate and thickness. In the Europan case for an estimated ocean salinity of 50~ppt, and using the constants from \citet{maykut1971}, these effects of salinity on the effective conductivity and heat capacity are most relevant where the shell is within about 10~K of the freezing temperature. To consider how mushy layer physics (and local fluid convection associated with brine rejection, see e.g., \citet{wettlaufer1997}) may interface with an overlying pure shell itself undergoing solid-state convection in the bulk, it may be necessary to design a coupled model which considers both the lower two-phase shell closest to the freezing point and the possibly Rayleigh-unstable overlying pure shell.

A method to use ice thickness and temperature measurements to infer turbulent ice-ocean heat fluxes related to a growing or melting Arctic sea ice floe has been developed previously by \citet{wettlaufer1991} and \citet{wettlaufer1990}. Here, we expect that observations of steady-state Europan ice thicknesses may be helpful at inferring the subsurface vertical ocean velocities, via inferences of the ice-ocean heat flux. The ice-ocean heat flux describes the rate at which heat is transported from the ocean to the base of the ice shell. Heat from the ocean may reach the shell either via molecular diffusion or turbulent convection; we expect the larger quantity to be convective transport. (See, e.g., \citet{ashkenazy2021} who describe a turbulent vertical flow.) In this case the ice-ocean heat flux ($F_w$) can be described as $F_w = \rho_o c_{p,o}\overline{w'T'}$, where $\rho_o$ is the density of the ocean, $c_{p,o}$ is the specific heat capacity of the ocean, $w'$ is a turbulent fluctuation to a background vertical ocean velocity, and $T'$ is a turbulent fluctuation to the background ocean temperature. 

If the viscosity is large enough and the shell is in a conductive steady state (with a known rate of internal heating, and possibly negligible for large viscosities in the depth-dependent prescription), an observation of ice thickness would give insight into the ice-ocean heat fluxes at the base of the ice shell, which can then be related to possible vertical (upwelling) ocean movements. In a conductive steady state, the ice thickness can be directly related to the ice-ocean heat flux and rate of internal heating (Equation \ref{eq:stef_steady}). This means that if a hypothetical steady-state conductive ice thickness (with a known rate of internal heating) were measured, one could directly calculate the ice-ocean heat flux, and thus $\overline{w'T'}$. If each temperature perturbation $T'$ has a magnitude $\delta T$, then the average value of the perturbation to the vertical velocity is $\overline{w'}=(F_w)/(\rho_o c_{p,o} \delta T)$. If $\delta T\sim10^{-3}-10^{-4}$~$^\circ$C \citep[inferred from Figure 4 of][]{ashkenazy2021}, $\rho_o\sim1000$~kg~m$^{-3}$, and $c_{p,o}\sim4000$~J~kg$^{-1}$~K$^{-1}$, then for the range of $F_w$ considered here (0.01~Wm$^{-2}$ to 0.1~Wm$^{-2}$), $\overline{w'}$ would vary between O(10$^{-3}$)~mm s$^{-1}$ to O(10$^{-1}$)~mm s$^{-1}$. Note, however, that, in general, $\overline{w'T'} \neq \overline{w'}~\overline{T'}$. If an order-of-magnitude estimate for the perturbation to temperature $T'$ can be attained, the above method provides one crude method to infer order-of-magnitude estimates for turbulent ocean velocities from an observation of ice thickness. 

Given the impending launch of Europa \textit{Clipper}, the REASON ice-penetrating radar measurements may help to constrain not only Europa's ice shell thickness, but also ice-ocean heat fluxes and vertical ocean velocities. The success of these measurements depends on the thermal profile of the ice and any impurities or liquid water present within; it is expected that a conductive, pure shell will be the most successful medium for an interpretable return signal \citep[e.g.,][]{kalousova2017}. If the shell is conductive and pure, our expected thicknesses (between about 4 to 40~km) should be largely penetrable via the radar \citep[see][who consider thicknesses up to 30~km]{kalousova2017}. This suggests that if the radar both penetrates to the bottom of the ice-ocean interface and returns a clear signal, the shell will be likely be conductive; in this way, it may be possible to constrain a lower bound on the basal viscosity and ice-ocean heat flux that would allow such a scenario. It may then be possible to make inferences about subsurface ocean dynamics by considering ice-ocean heat fluxes from the thermodynamic model described herein.

\clearpage

\noindent NCS acknowledges support from the Princeton Center for Theoretical Science. NCS acknowledges helpful conversations with Ching-Yao Lai.


\begin{thebibliography}{}
\expandafter\ifx\csname natexlab\endcsname\relax\def\natexlab#1{#1}\fi
\providecommand{\url}[1]{\href{#1}{#1}}
\providecommand{\dodoi}[1]{doi:~\href{http://doi.org/#1}{\nolinkurl{#1}}}
\providecommand{\doeprint}[1]{\href{http://ascl.net/#1}{\nolinkurl{http://ascl.net/#1}}}
\providecommand{\doarXiv}[1]{\href{https://arxiv.org/abs/#1}{\nolinkurl{https://arxiv.org/abs/#1}}}

\bibitem[{{Allu Peddinti} \& McNamara(2019)}]{peddinti2019}
{Allu Peddinti}, D., \& McNamara, A.~K. 2019, Icarus, 329, 251, \dodoi{10.1016/j.icarus.2019.03.037}

\bibitem[{Ashkenazy(2019)}]{ashkenazy2019}
Ashkenazy, Y. 2019, Heliyon, 5, e01908, \dodoi{10.1016/j.heliyon.2019.e01908}

\bibitem[{Ashkenazy {et~al.}(2018)Ashkenazy, Sayag, \& Tziperman}]{ashkenazy2018}
Ashkenazy, Y., Sayag, R., \& Tziperman, E. 2018, Nature Astronomy, 2, 43, \dodoi{10.1038/s41550-017-0326-7}

\bibitem[{Ashkenazy \& Tziperman(2021)}]{ashkenazy2021}
Ashkenazy, Y., \& Tziperman, E. 2021, Nature communications, 12, 6376, \dodoi{10.1038/s41467-021-26710-0}

\bibitem[{Barr \& McKinnon(2007)}]{barr2007}
Barr, A.~C., \& McKinnon, W.~B. 2007, Journal of Geophysical Research: Planets, 112, E02012, \dodoi{10.1029/2006JE002781}

\bibitem[{Barr \& Pappalardo(2005)}]{barr2005}
Barr, A.~C., \& Pappalardo, R.~T. 2005, Journal of Geophysical Research: Planets, 110, E12005, \dodoi{10.1029/2004JE002371}

\bibitem[{Barr {et~al.}(2004)Barr, Pappalardo, \& Zhong}]{barr2004}
Barr, A.~C., Pappalardo, R.~T., \& Zhong, S. 2004, Journal of Geophysical Research: Planets, 109, E12008, \dodoi{10.1029/2004JE002296}

\bibitem[{B{\'e}nard(1900)}]{benard1900}
B{\'e}nard, H. 1900, Rev. Gen. Sci. Pures Appl., 11, 1261

\bibitem[{{Bierson} \& {Nimmo}(2020)}]{bierson2020}
{Bierson}, C.~J., \& {Nimmo}, F. 2020, \apjl, 897, L43, \dodoi{10.3847/2041-8213/aba11a}

\bibitem[{Buffo {et~al.}(2021{\natexlab{a}})Buffo, Schmidt, Huber, \& Meyer}]{buffo2021}
Buffo, J., Schmidt, B., Huber, C., \& Meyer, C. 2021{\natexlab{a}}, Icarus, 360, 114318, \dodoi{https://doi.org/10.1016/j.icarus.2021.114318}

\bibitem[{Buffo {et~al.}(2021{\natexlab{b}})Buffo, Meyer, \& Parkinson}]{buffo2021b}
Buffo, J.~J., Meyer, C.~R., \& Parkinson, J. R.~G. 2021{\natexlab{b}}, Journal of Geophysical Research: Planets, 126, e2020JE006741, \dodoi{https://doi.org/10.1029/2020JE006741}

\bibitem[{Buratti \& Veverka(1983)}]{buratti1983}
Buratti, B., \& Veverka, J. 1983, Icarus, 55, 93, \dodoi{10.1016/0019-1035(83)90053-2}

\bibitem[{{Burns} {et~al.}(2020){Burns}, {Vasil}, {Oishi}, {Lecoanet}, \& {Brown}}]{Burns+2020}
{Burns}, K.~J., {Vasil}, G.~M., {Oishi}, J.~S., {Lecoanet}, D., \& {Brown}, B.~P. 2020, Physical Review Research, 2, 023068, \dodoi{10.1103/PhysRevResearch.2.023068}

\bibitem[{{Carnahan} {et~al.}(2021){Carnahan}, {Wolfenbarger}, {Jordan}, \& {Hesse}}]{Carnahan+2021}
{Carnahan}, E., {Wolfenbarger}, N.~S., {Jordan}, J.~S., \& {Hesse}, M.~A. 2021, Earth and Planetary Science Letters, 563, 116886, \dodoi{10.1016/j.epsl.2021.116886}

\bibitem[{Carr {et~al.}(1998)Carr, Belton, Chapman, Davies, Geissler, Greenberg, McEwen, Tufts, Greeley, Sullivan, {et~al.}}]{carr1998}
Carr, M.~H., Belton, M.~J., Chapman, C.~R., {et~al.} 1998, Nature, 391, 363, \dodoi{10.1038/34857}

\bibitem[{Cassen {et~al.}(1980)Cassen, Peale, \& Reynolds}]{cassen1980}
Cassen, P., Peale, S.~J., \& Reynolds, R.~T. 1980, Geophysical Research Letters, 7, 987, \dodoi{10.1029/GL007i011p00987}

\bibitem[{{Cassen} {et~al.}(1979){Cassen}, {Reynolds}, \& {Peale}}]{Cassen+1979}
{Cassen}, P., {Reynolds}, R.~T., \& {Peale}, S.~J. 1979, \grl, 6, 731, \dodoi{10.1029/GL006i009p00731}

\bibitem[{Durham \& Stern(2001)}]{durham2001}
Durham, W., \& Stern, L. 2001, Annual Review of Earth and Planetary Sciences, 29, 295, \dodoi{10.1146/annurev.earth.29.1.295}

\bibitem[{Feltham {et~al.}(2006)Feltham, Untersteiner, Wettlaufer, \& Worster}]{feltham2006}
Feltham, D.~L., Untersteiner, N., Wettlaufer, J.~S., \& Worster, M.~G. 2006, Geophysical Research Letters, 33, L14501, \dodoi{https://doi.org/10.1029/2006GL026290}

\bibitem[{Glen(1955)}]{glen1955}
Glen, J.~W. 1955, Proceedings of the Royal Society of London. Series A, Mathematical and Physical Sciences, 228, 519, \dodoi{10.1098/rspa.1955.0066}

\bibitem[{{Goldsby} \& {Kohlstedt}(2001)}]{Goldsby+Kohlstedt2001}
{Goldsby}, D.~L., \& {Kohlstedt}, D.~L. 2001, \jgr, 106, 11,017, \dodoi{10.1029/2000JB900336}

\bibitem[{Green {et~al.}(2021)Green, Montesi, \& Cooper}]{green2021}
Green, A., Montesi, L., \& Cooper, C. 2021, Journal of Geophysical Research: Planets, 126, e2020JE006677, \dodoi{10.1029/2020JE006677}

\bibitem[{Greenberg {et~al.}(2000)Greenberg, Geissler, Tufts, \& Hoppa}]{greenberg2000}
Greenberg, R., Geissler, P., Tufts, B.~R., \& Hoppa, G.~V. 2000, Journal of Geophysical Research: Planets, 105, 17551, \dodoi{https://doi.org/10.1029/1999JE001147}

\bibitem[{Hammond {et~al.}(2018)Hammond, Parmenteir, \& Barr}]{hammond2018}
Hammond, N.~P., Parmenteir, E.~M., \& Barr, A.~C. 2018, Journal of Geophysical Research: Planets, 123, 3105, \dodoi{https://doi.org/10.1029/2018JE005781}

\bibitem[{Hand \& Chyba(2007)}]{hand2007}
Hand, K.~P., \& Chyba, C.~F. 2007, Icarus, 189, 424 , \dodoi{10.1016/j.icarus.2007.02.002}

\bibitem[{Hand {et~al.}(2009)Hand, Chyba, Priscu, Carlson, \& Nealson}]{hand2009}
Hand, K.~P., Chyba, C.~F., Priscu, J.~C., Carlson, R.~W., \& Nealson, K.~H. 2009, in Europa, ed. R.~T. Pappalardo, W.~B. McKinnon, \& K.~Khurana (University of Arizona Press), 589--629

\bibitem[{{Herring}(1950)}]{Herring1950}
{Herring}, C. 1950, Journal of Applied Physics, 21, 437, \dodoi{10.1063/1.1699681}

\bibitem[{Hoppa {et~al.}(1999)Hoppa, Tufts, Greenberg, \& Geissler}]{hoppa1999}
Hoppa, G.~V., Tufts, B.~R., Greenberg, R., \& Geissler, P.~E. 1999, Science, 285, 1899, \dodoi{10.1126/science.285.5435.1899}

\bibitem[{Howell(2021)}]{howell2021}
Howell, S.~M. 2021, The Planetary Science Journal, 2, 129, \dodoi{10.3847/psj/abfe10}

\bibitem[{Hussmann \& Spohn(2004)}]{hussmann2004}
Hussmann, H., \& Spohn, T. 2004, Icarus, 171, 391, \dodoi{10.1016/j.icarus.2004.05.020}

\bibitem[{Hussmann {et~al.}(2002)Hussmann, Spohn, \& Wieczerkowski}]{hussmann2002}
Hussmann, H., Spohn, T., \& Wieczerkowski, K. 2002, Icarus, 156, 143, \dodoi{10.1006/icar.2001.6776}

\bibitem[{Jacka \& Jun(1994)}]{jacka1994steady}
Jacka, T., \& Jun, L. 1994, Annals of Glaciology, 20, 13

\bibitem[{Jain \& Solomatov(2022)}]{jain2022}
Jain, C., \& Solomatov, V.~S. 2022, Physics of Fluids, 34, 096604, \dodoi{10.1063/5.0105170}

\bibitem[{Kalousová {et~al.}(2017)Kalousová, Schroeder, \& Soderlund}]{kalousova2017}
Kalousová, K., Schroeder, D.~M., \& Soderlund, K.~M. 2017, Journal of Geophysical Research: Planets, 122, 524, \dodoi{https://doi.org/10.1002/2016JE005110}

\bibitem[{Khurana {et~al.}(2002)Khurana, Kivelson, \& Russell}]{khurana2002}
Khurana, K.~K., Kivelson, M.~G., \& Russell, C.~T. 2002, Astrobiology, 2, 93, \dodoi{10.1089/153110702753621376}

\bibitem[{Kirk \& Stevenson(1987)}]{kirk1987}
Kirk, R., \& Stevenson, D. 1987, Icarus, 69, 91, \dodoi{https://doi.org/10.1016/0019-1035(87)90009-1}

\bibitem[{Kivelson {et~al.}(2000)Kivelson, Khurana, Russell, Volwerk, Walker, \& Zimmer}]{kivelson2000}
Kivelson, M.~G., Khurana, K.~K., Russell, C.~T., {et~al.} 2000, Science, 289, 1340, \dodoi{10.1126/science.289.5483.1340}

\bibitem[{Lewis(1971)}]{lewis1971}
Lewis, J.~S. 1971, Icarus, 15, 174, \dodoi{10.1016/0019-1035(71)90072-8}

\bibitem[{Maykut \& Untersteiner(1971)}]{maykut1971}
Maykut, G.~A., \& Untersteiner, N. 1971, Journal of Geophysical Research, 76, 1550, \dodoi{https://doi.org/10.1029/JC076i006p01550}

\bibitem[{McKinnon(1999)}]{mckinnon1999}
McKinnon, W.~B. 1999, Geophysical Research Letters, 26, 951, \dodoi{10.1029/1999GL900125}

\bibitem[{Mitri \& Showman(2005)}]{mitri2005}
Mitri, G., \& Showman, A.~P. 2005, Icarus, 177, 447, \dodoi{10.1016/j.icarus.2005.03.019}

\bibitem[{Moore(2006)}]{moore2006}
Moore, W.~B. 2006, Icarus, 180, 141, \dodoi{10.1016/j.icarus.2005.09.005}

\bibitem[{Nimmo {et~al.}(2007)Nimmo, Thomas, Pappalardo, \& Moore}]{nimmo2007}
Nimmo, F., Thomas, P., Pappalardo, R., \& Moore, W. 2007, Icarus, 191, 183, \dodoi{10.1016/j.icarus.2007.04.021}

\bibitem[{Ojakangas \& Stevenson(1989)}]{ojakangas1989}
Ojakangas, G.~W., \& Stevenson, D.~J. 1989, Icarus, 81, 220, \dodoi{10.1016/0019-1035(89)90052-3}

\bibitem[{Pappalardo {et~al.}(1998)Pappalardo, Head, Greeley, Sullivan, Pilcher, Schubert, Moore, Carr, Moore, Belton, {et~al.}}]{pappalardo1998}
Pappalardo, R., Head, J., Greeley, R., {et~al.} 1998, Nature, 391, 365, \dodoi{10.1038/34862}

\bibitem[{Pappalardo {et~al.}(1999)Pappalardo, Belton, Breneman, Carr, Chapman, Collins, Denk, Fagents, Geissler, Giese, Greeley, Greenberg, Head, Helfenstein, Hoppa, Kadel, Klaasen, Klemaszewski, Magee, McEwen, Moore, Moore, Neukum, Phillips, Prockter, Schubert, Senske, Sullivan, Tufts, Turtle, Wagner, \& Williams}]{pappalardo1999}
Pappalardo, R.~T., Belton, M. J.~S., Breneman, H.~H., {et~al.} 1999, Journal of Geophysical Research: Planets, 104, 24015, \dodoi{https://doi.org/10.1029/1998JE000628}

\bibitem[{Perovich \& Polashenski(2012)}]{perovich2012}
Perovich, D.~K., \& Polashenski, C. 2012, Geophysical Research Letters, 39, L08501, \dodoi{10.1029/2012GL051432}

\bibitem[{Quick \& Marsh(2015)}]{quick2015}
Quick, L.~C., \& Marsh, B.~D. 2015, Icarus, 253, 16, \dodoi{10.1016/j.icarus.2015.02.016}

\bibitem[{Rayleigh(1916)}]{rayleigh1916}
Rayleigh, L. 1916, The London, Edinburgh, and Dublin Philosophical Magazine and Journal of Science, 32, 529, \dodoi{10.1080/14786441608635602}

\bibitem[{Shibley \& Laughlin(2021)}]{shibley2021}
Shibley, N.~C., \& Laughlin, G. 2021, The Planetary Science Journal, 2, 221, \dodoi{10.3847/PSJ/ac2b2c}

\bibitem[{Showman \& Han(2004)}]{showman2004}
Showman, A.~P., \& Han, L. 2004, Journal of Geophysical Research: Planets, 109, \dodoi{10.1029/2003JE002103}

\bibitem[{Solomatov(1995)}]{solomatov1995}
Solomatov, V.~S. 1995, Physics of Fluids, 7, 266, \dodoi{10.1063/1.868624}

\bibitem[{Steinbrugge {et~al.}(2020)Steinbrugge, Voigt, Wolfenbarger, Hamilton, Soderlund, Young, Blankenship, Vance, \& Schroeder}]{steinbrugge2020}
Steinbrugge, G., Voigt, J. R.~C., Wolfenbarger, N.~S., {et~al.} 2020, Geophysical Research Letters, 47, e2020GL090797, \dodoi{https://doi.org/10.1029/2020GL090797}

\bibitem[{Stengel {et~al.}(1982)Stengel, Oliver, \& Booker}]{stengel1982}
Stengel, K.~C., Oliver, D.~S., \& Booker, J.~R. 1982, Journal of Fluid Mechanics, 120, 411–431, \dodoi{10.1017/S0022112082002821}

\bibitem[{Thomson \& Delaney(2001)}]{thompson2001}
Thomson, R.~E., \& Delaney, J.~R. 2001, Journal of Geophysical Research: Planets, 106, 12355, \dodoi{10.1029/2000JE001332}

\bibitem[{Tobie {et~al.}(2003)Tobie, Choblet, \& Sotin}]{tobie2003}
Tobie, G., Choblet, G., \& Sotin, C. 2003, Journal of Geophysical Research: Planets, 108, \dodoi{10.1029/2003JE002099}

\bibitem[{Turtle \& Pierazzo(2001)}]{turtle2001}
Turtle, E.~P., \& Pierazzo, E. 2001, Science, 294, 1326, \dodoi{10.1126/science.1062492}

\bibitem[{Vance {et~al.}(2021)Vance, Styczinski, Bills, Cochrane, Soderlund, Gómez-Pérez, \& Paty}]{vance2021}
Vance, S.~D., Styczinski, M.~J., Bills, B.~G., {et~al.} 2021, Journal of Geophysical Research: Planets, 126, e2020JE006418, \dodoi{https://doi.org/10.1029/2020JE006418}

\bibitem[{Weertman(1983)}]{weertman1983creep}
Weertman, J. 1983, Annual Review of Earth and Planetary Sciences, 11, 215

\bibitem[{Wettlaufer {et~al.}(1990)Wettlaufer, Untersteiner, \& Colony}]{wettlaufer1990}
Wettlaufer, J., Untersteiner, N., \& Colony, R. 1990, Annals of Glaciology, 14, 315–318, \dodoi{10.3189/S026030550000882X}

\bibitem[{Wettlaufer {et~al.}(1997)Wettlaufer, Worster, \& Huppert}]{wettlaufer1997}
Wettlaufer, J., Worster, M.~G., \& Huppert, H.~E. 1997, Journal of Fluid Mechanics, 344, 291–316, \dodoi{10.1017/S0022112097006022}

\bibitem[{Wettlaufer(1991)}]{wettlaufer1991}
Wettlaufer, J.~S. 1991, Journal of Geophysical Research: Oceans, 96, \dodoi{10.1029/90JC00081}

\bibitem[{Zolotov \& Shock(2001)}]{zolotov2001}
Zolotov, M.~Y., \& Shock, E.~L. 2001, Journal of Geophysical Research: Planets, 106, 32815, \dodoi{10.1029/2000JE001413}

\end{thebibliography}
\end{document}